\documentclass[useAMS,usenatbib]{mn2e}
\usepackage{graphics,epsfig,aas_macros}
\usepackage{graphicx}
\usepackage[normalem]{ulem}
\usepackage[dvipsnames]{color}
\usepackage[]{inputenc,amssymb}
\usepackage{amsmath}
\usepackage{color}
\usepackage{textcomp}
\usepackage{array}
\usepackage{textcase}
\usepackage{titlesec}
\usepackage{etoolbox}
\usepackage{comment}

\makeatletter
\patchcmd{\ttlh@hang}{\parindent\z@}{\parindent\z@\leavevmode}{}{}
\patchcmd{\ttlh@hang}{\noindent}{}{}{}
\makeatother
\renewcommand\footnoterule{\kern 5pt \hrule width 2in \kern 5.0pt}

\titleformat{\paragraph}
{\it\normalsize\normalsize}{\theparagraph}{1em}{}
\titlespacing*{\paragraph}
{0pt}{3.25ex plus 1ex minus .2ex}{1.5ex plus .2ex}

\newcolumntype{P}[1]{>{\centering\arraybackslash}p{#1}}

\usepackage[toc,page]{appendix}

\def \be{\begin{equation}}
\def \ee{\end{equation}}
\def \bea{\begin{eqnarray}}
\def \eea{\end{eqnarray}}

\def \fg{$f_{\rm gas}$ }

\title[Hot CGM in massive galaxies]{X-ray and SZ constraints on the properties of hot CGM}

\author [Singh, Majumdar, Nath and Silk]
{Priyanka Singh$^{1}$ \thanks{priyankas@iucaa.in}, 
Subhabrata Majumdar$^2$, \thanks{subha@tifr.res.in},
Biman B. Nath$^3$ % \thanks{biman@rri.res.in}, 
\& Joseph Silk$^{4,5,6}$ %\thanks{j.silk1@physics.ox.ac.uk} 
 \\
$^1$ Inter-University Centre for Astronomy and Astrophysics, Ganeshkhind, Post Bag 4, Pune 411007, India\\
$^2$ Tata Institute of Fundamental Research, Mumbai, India, 400005\\
$^3$ Raman Research Institute, Bangalore, India, 560080 \\
$^{4}$ Institut d'Astrophysique de Paris (UMR 7095: CNRS \& UPMC -- Sorbonne
Universit\'es), 98 bis bd Arago, F-75014 Paris, France\\ 
$^{5}$ Department of Physics and Astronomy, The Johns Hopkins University Homewood Campus, Baltimore, MD 21218, USA\\
$^{6}$ BIPAC, Department of Physics, University of Oxford, Keble Road, Oxford OX1 3RH, UK
}

\begin{document}
\label{firstpage}
% \pagerange{\pageref{firstpage}--\pageref{lastpage}}
\maketitle

\begin{abstract}
We use observations of stacked X-ray luminosity and Sunyaev-Zel'dovich (SZ) signal from a
cosmological sample of $\sim 80,000$ and $104,000$ massive galaxies, respectively, with 
$ 10^{12.6}\lesssim M_{500} \lesssim 10^{13} M_{\odot}$ and mean redshift, \={z} $\sim$ 0.1 - 0.14 to
constrain
the hot Circumgalactic Medium (CGM) density and temperature. The X-ray luminosities
constrain the density and hot CGM mass, while the SZ signal
helps in breaking the density-temperature degeneracy.
We consider a simple power-law density 
distribution ($n_e \propto r^{-3\beta}$) as well as a
hydrostatic hot halo model, with the gas assumed to be isothermal in both cases.
The datasets are best described by the mean hot CGM profile $\propto r^{-1.2}$, which is shallower than an NFW profile.
For halo virial mass $\sim 10^{12}$ - $10^{13} M_{\odot}$, the hot CGM contains
$\sim$ 20 - 30\% of galactic baryonic mass for the power-law model
and 4 - 11\% for the hydrostatic halo model, within the virial radii. 
For the power-law model, the hot CGM profile broadly agrees with observations of the Milky Way.
The mean hot CGM mass is comparable to or larger than the mass contained in other phases of the 
CGM for $L^*$ galaxies. 
\end{abstract}


\section{Introduction}
The evidence for the absence of a significant fraction of baryons, 
otherwise predicted by the $\Lambda$-CDM cosmological model, covers a large number of 
observations (eg. \citealt{klypin02, bell03, flynn06, anderson10, mcgaugh10, miller15}). 
This lack of the baryons, also known as the 
\textquotedblleft galactic missing baryons problem\textquotedblright \,  becomes
severe as one goes down to the range of galactic masses.
Analytical as well as numerical studies have indicated the existence of a
hot (T $>10^6$ K) and diffuse circumgalactic medium (CGM), 
occupying the galactic halo \citep{rees77, silk77, white91, maller04, keres05, crain10, sharma12}.
Recent observations of the Milky Way and other massive galaxies also support the presence of a
significant fraction of galactic baryons in the hot phase of CGM
(eg. \citealt{putman09, anderson11, dai12, gatto13}). 
Previous estimates from stacking of $\sim 2000$ galaxies yielded hot CGM masses of 
$1.5-3.3\times10^{10} M_\odot$ within 200 kpc \citep{anderson13}.
The hot CGM is also predicted to be detectable through its SZ and X-ray power spectra
using the combination of high resolution surveys such as South Pole Telescope,
the extended ROentgen Survey with an Imaging Telescope Array and the Dark Energy Survey \citep{singh15a, singh15b}.
Other phases of CGM such as warm (T $\sim 10^5-10^6$ K), cool (T $\sim 10^4-10^5$ K)
and the cold (T $< 10^4$ K) phases
are also expected to contribute significantly to the total amount of CGM.
However, there is considerable uncertainty in the knowledge of 
total CGM mass, contribution of different CGM phases
and their density profiles (eg. \citealt{tumlinson17}).

\begin{table*}
\caption{Summary of tSZ and X-ray datasets used here.}
\centering
\resizebox{0.65\textwidth}{!}{
\setlength{\tabcolsep}{3pt}
\begin{tabular}{c c c c c c}
\\
\hline \\
$\log M_* $ & $\log M_{500} (M_{\odot})$ & $\tilde{Y}_{500} \pm \sigma \tilde{Y}_{500}$ 
& $\log L^{\rm CGM} _X$ $\pm$ $\sigma L^{\rm CGM} _X$ & \={z}&  Number of LBGs stacked\\
$M_{\odot}$ & P13(A15) & $\rm 10^{-6} arcmin^2$ & $\log$ ergs/s & & P13(A15)\\\\
\hline\\
11.15 & 12.97 (13.09) & 1.7 $\pm$ 1.0 & 40.99 $\pm$ 0.11 & 0.135 & 22085 (18430)\\\\
11.05 & 12.71 (12.91) & 1.27 $\pm$ 0.78 & 40.55 $\pm$ 0.53 & 0.127 & 26026 (21583)\\\\
10.95 & 12.62 (12.75) & 1.54 $\pm$ 0.60 & 40.28 $\pm$ 0.48 & 0.113 & 28325 (22689)\\\\
10.85 & 12.40 (12.60) & - & 39.28 $\pm$ 0.93 & 0.105 & 27866 (22490)\\\\
 \hline
 \end{tabular}
 \label{tab-data}}
\end{table*}

In this paper, we constrain the properties of hot CGM 
using the stacked soft (0.5-2 keV) X-ray emission 
detected down to $M_* \sim 6\times 10^{10} M_{\odot}$ by \cite{anderson15} (hereafter, A15)
and stacked thermal SZ (tSZ) signal detected down to $M_* \sim 10^{11} M_{\odot}$ by the \cite{planck13} 
(hereafter, P13).
We aim to obtain a simple analytical model of 
the hot gas distribution for galaxies ($M_v \sim 10^{12}-10^{13}
M_{\odot}$), which can explain the above measurements consistently.
To our knowledge, it is the first study combining stacked tSZ and X-ray measurements to 
constrain CGM properties focused on the galaxy mass regime.
We do not attempt to fit the full mass range observed by P13 and A15, whose data 
include those of galaxy clusters and groups.
A single characterization of hot gas 
is not expected in such different classes of objects.

{\section{Datasets}
In this section, we describe the datasets used here and the physical
processes underlying tSZ effect and X-ray emission.
\subsection{Thermal SZ effect}
\label{sec-tsz-data}
It has been difficult to detect tSZ signal from galaxies due to their small gas reservoir.
However, this situation 
can be improved upon by stacking a large number of galaxies thus increasing the signal-to-noise ratio (SNR).

P13 stacked {\it Planck} tSZ signal from a large number ($\sim 2.5\times 10^5$) of
locally brightest galaxies (LBGs), divided into twenty logarithmically equally spaced stellar mass bins.
They detected the stacked signal with SNR $> 3 \sigma$
at $M_* > 2\times 10^{11} M_{\odot}$ ($M_{500}
\footnote {Mass enclosed within a radius $R_{500}$
such that the mean density is 500 times the critical density of the Universe.}
>2\times 10^{13} M_{\odot}$), whereas, the stacked signal is marginally detected
(SNR $\sim 1.6$ to $2.6 \sigma$) down to $M_{*} \sim 10^{11} M_{\odot}$ ($M_{500} \sim 4\times 10^{12} M_{\odot}$).
The LBG sample is obtained after applying a series of selection criteria
on New York University Value Added Galaxy Catalogue based 
on SDSS-DR7 (see P13 for details of the selection criteria). 
The selection criteria ensure that each galaxy in the sample is central to its 
dark matter halo. The stellar mass of each LBG is obtained from SDSS 
photometry \citep{blanton07}. In order to connect the stellar mass to the host dark matter halo properties, 
P13 made use of a mock galaxy catalogue created by Millennium Simulation which is tuned to mimic the 
SDSS galaxy catalogue \citep{ springel05b, guo13}. 

The stacked tSZ signal, $Y_{500}$
is the Compton y-parameter integrated over the sphere of radius $R_{500}$,
\begin{equation}
Y_{500}=\frac{\sigma_T}{m_e c^2 D^2 _A (z)} \int_0 ^{R_{500}} P_e dV, 
\end{equation}
where $D_A$ is the angular diameter distance, $P_e=n_e k_{\rm b} T_e$ is the electron 
pressure, $n_e$ and $T_e$ are electron density and temperature, respectively.
Instead of dealing directly with $Y_{500}$, the results are shown in terms of $\tilde{Y}_{500}$, which is the 
tSZ signal scaled to $z=0$ and to a fixed angular diameter distance. It is related to $Y_{500}$ as
\begin{equation}
 \tilde{Y}_{500} \equiv Y_{500} E^{-2/3}(z) (D_A(z)/500 \rm Mpc)^2
\end{equation}

 P13 do not directly measure 
$\tilde{Y}_{500}$ due to the large beam size of {\it Planck}. 
Instead, they measure cylindrically integrated tSZ signal within a much
larger aperture of size $5 R_{500}$, which is given by,

\begin{equation}
 Y_{\rm cyl}=\frac{\sigma_T}{m_e c^2 D^2 _A(z)}  
 \int_0 ^{5 R_{500}} 2\pi r dr \int_r ^{5 R_{500}} \frac{2 P_e(r')r'
 dr'}{\sqrt{r'^2-r^2}}
 \label{eqn-ycyl}
\end{equation}

The cylindrical tSZ signal, $Y_{\rm cyl}$ is then converted into $Y_{500}$ assuming a pressure profile of the gas.
P13 assume that the gas follows Universal pressure profile \citep{arnaud10} to convert $Y_{\rm cyl}$ to $Y_{500}$.
The conversion factor ($Y_{\rm cyl}/Y_{500}$) is close to two for the Universal pressure profile \citep{brun15, greco15}.
However, this conversion factor may vary significantly for different pressure profiles and 
halo masses \citep{greco15}.
Therefore, it is appropriate to compare the results for other gas distributions with cylindrical tSZ signal.
In this paper, we use Equation \ref{eqn-ycyl} to compute $\tilde{Y}_{\rm cyl}$ for other pressure profiles
and then compare our predictions directly with the measurements of cylindrical tSZ signal.

\subsection{X-ray emission}
The hot phase of CGM also manifests itself in X-rays due to its high temperature.
A15 stacked X-ray luminosity of LBGs in the soft X-ray band 
(0.5-2 keV) of ROSAT all sky survey, thus detecting X-ray emission from the hot gas down
to $M_* \sim 10^{10.8} M_{\odot}$. Additionally, they measured X-ray emission arising only from region 
$(0.15-1)\times R_{500}$, referring to it as the circumgalactic emission. 
They start with the same sample of LBGs as used by P13 and apply additional selection criteria
(see section 3.1 and Figure 1 of A15) thus producing a slightly smaller sample of LBGs.
A15 estimated the effective halo mass, $M_{500}$ for twelve highest stellar mass bins of the sample using
their best-fitting $L_X-M_{500}$ relation, whereas P13 used their $Y_{500}-M_{500}$ relation with \cite{arnaud10} pressure
profile to get $M_{500}$. The difference between the two is small.
The LBGs span a redshift range $\sim 0.1-0.14$ in the mass range of interest.
We summarize both datasets in the mass range of our interest in Table \ref{tab-data}.
The uncertainties quoted in the table (and used in this work) are bootstrap errors and the mean redshift
is computed from the mean luminosity distances. We use {\it WMAP7} cosmology throughout this paper. 
 
Analytically, the X-ray luminosity of the hot CGM ($0.15R_{500}$ to $R_{500}$) can be computed
using the following relation,
\begin{equation}
 L^{\rm CGM} _X = \int_{0.15 R_{500}} ^{R_{500}}
 2\pi r dr \int_r ^{R_{500}} \frac{2\, n_e n_i \Lambda(Z,T_e) r'
 dr'}{\sqrt{r'^2-r^2}},
\end{equation}
where, $Z$ is the CGM metallicity, $n_i$ is the proton density and
$\Lambda(Z,T_e)$ is the cooling function. We use the 
Astrophysical Plasma Emission Code (APEC; \citealt{smith01}) to calculate $\Lambda(Z,T)$.
We fix the CGM metallicity at $Z=0.2$ \citep{li17} for our main results
and explore the effects of a different metallicity in section \ref{sec-error}.
Similar to tSZ results, we scale the X-ray luminosity to $z=0$, denoted by, 
$\tilde{L}^{\rm CGM} _X=L^{\rm CGM} _X E^{-7/3}(z)$. 

\begin{figure}
% \begin{center}
\includegraphics[height=7.0cm,width=7.8cm,angle=0.0 ]{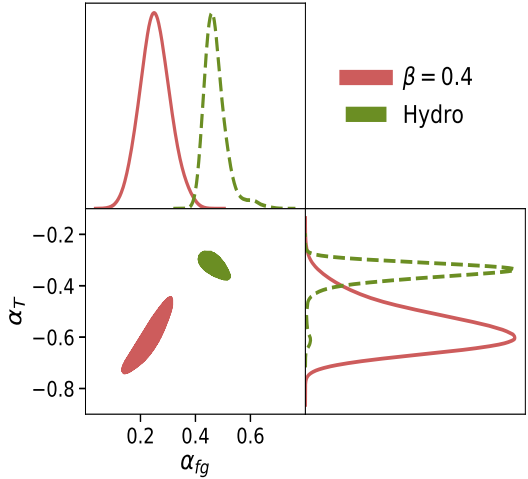}
% \end{center}
\caption{The 68\% CL contours for $\alpha_{\rm fg}$
and $\alpha_{T}$ computed using MCMC. The solid (red) lines and contour correspond to the power-law model whereas
dashed (green) lines and contour corresponds to hydrostatic hot halo model.
}
\label{fig-mcmc}
\end{figure}

\section{CGM density and temperature}
We use Markov chain Monte Carlo (MCMC)\footnote{emcee; \citealt{emcee}} analysis to 
determine the CGM density and temperature.
We explore the following two spherical gas distributions.

\subsection{A power-law model}
First, we consider a simple power-law density profile given by $n_e (r) \propto r^{-3 \beta}$.
The power-law density profile is equivalent to a standard $\beta$-model at radii larger than the core radius.
The gas fraction \fg (i.e. the ratio of gas mass within the virial radius $R_v$ and 
the total halo virial mass $M_v$\footnote{We define the virial mass and virial
radius in terms of overdensity, $\Delta_c(z)=18\pi^2+82(\Omega_M(z)-1)-39(\Omega_M(z) -1)^2$.}) is given by,
\begin{equation}
f_{\rm gas}= \frac{4 \pi \mu_e m_p}{M_v} \int^{R_v} _0 dr\, r^2\, n_e(r),
\label{eqn-fgas}
\end{equation}

where $\mu_e (=1.36)$ is the mean molecular mass per electron.
We assume the gas to be isothermal (to keep the model simple and reduce the number of free parameters),
at temperature $f_T$ times the virial temperature of the halo, i.e.
$T_{\rm gas}=f_T\times T_{\rm vir}$, where,
\begin{equation}
 T_{\rm vir} = \frac{\mu \, m_p G M_v}{2k_b R_v}.
\end{equation}

\begin{table*}
\caption{X-ray-tSZ joint constraints from MCMC analysis.}
\centering
\resizebox{0.75\textwidth}{!}{
\begin{tabular}{c c c c c c}
\\
\hline \\
  & Model & $\rm \alpha_{fg}$ (Mean $\pm$ 68\% CL) &  $\rm \alpha_{T}$ (Mean $\pm$ 68\% CL) & \fg ($10^{12} M_{\odot}$)
  &  \fg ($10^{13} M_{\odot}$)\\\\
 \hline\\
& $\beta=0.4$ & $0.24 \pm 0.061$ & $-0.59^{+0.071} _{-0.12}$ & $3.2^{+1.7} _{-1.1}$ \% & $5.5^{+1.8} _{-1.4}$ \% \\\\
& Hydro & $0.48^{+0.027} _{-0.051}$ & $-0.33^{+0.052} _{-0.023}$ & $0.6^{+0.3} _{-0.1}$ \% & $1.8^{+0.5} _{-0.2}$ \% \\\\
 \hline
 \end{tabular}
}
 \label{tab-szxr}
\end{table*}

\begin{figure}
\vspace{2mm}
% \begin{center}
% \includegraphics[width=8cm,angle=0.0 ]{y500-szxr.pdf}
% \includegraphics[width=8cm,angle=0.0 ]{lxcgm-szxr.pdf}
\includegraphics[width=8.2cm,angle=0.0 ]{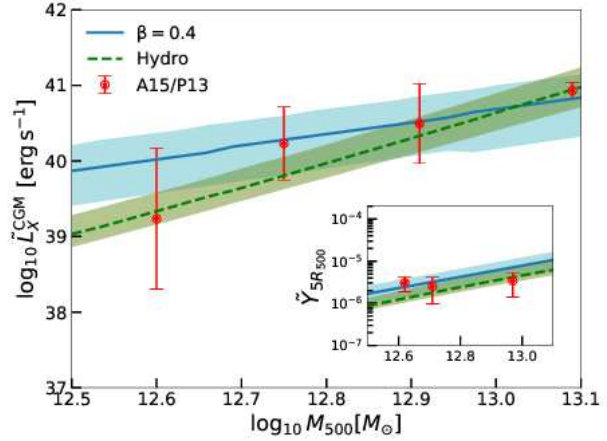}
% \end{center}
\caption{Stacked CGM soft X-ray luminosities and tSZ signal (inset)  compared to the
predictions of best-fitting parameter values (solid blue lines for power-law model and dashed green lines for 
hydrostatic halo model). The shaded regions signify 68\% CL regions.}
\label{fig-uncert}
\end{figure}

Here $\mu$ (=0.59, for primordial ionized gas) is the mean molecular weight of the gas.
We then use \fg and $f_T$ to define the free parameters of our model, namely $\alpha_{\rm fg}$
and $\alpha_{T}$, given by,
\begin{align}
  f_{\rm gas} &= \frac{\Omega_B}{\Omega_M} \Bigl(\frac{M_v}{10^{15} M_{\odot}} \Bigr)^{\alpha_{fg}} \nonumber \\
 f_T &= f_{12} \Bigl(\frac{M_v}{10^{12}  M_{\odot}} \Bigr)^{\alpha_T} \label{eqn-tv} 
\end{align}

The functional form of \fg is inspired from the 
observed deficit of hot gas in lower mass systems compared to the massive haloes \citep{bell03, mcgaugh10},
with the hot gas mass being close to the cosmic baryon fraction in the clusters. 
We use MCMC analysis to constrain $\alpha_{\rm fg}$ and $\alpha_{T}$.
For both parameters, we use uniform priors large enough that they do not 
affect the results of the fitting process.
GetDIst python package\footnote{http://getdist.readthedocs.io/en/latest/index.html}
is used to analyze and plot the results of MCMC analysis.

There are two more free parameters $f_{12}$ (the value of $f_T$ at $M_v=10^{12} M_{\odot}$) 
and $\beta$ in the formalism described above.
The observed temperature of the hot gas in the Milky Way \citep{miller15}
and external galaxies with $M_v \gtrsim 10^{12} M_{\odot}$ \citep{li17}
is generally $\rm \gtrsim 2\times 10^{6} K$.
Therefore, we fix $f_{12}\sim $3.4 (for the above mentioned definition of virial temperature).
We find that the reduced-$\chi^2\rightarrow 1$ 
for $\beta=0.4$. A flatter gas distribution ($\beta < 0.4$) gives a bad fit to the data (reduced-$\chi^2 > 1$), 
whereas a steeper gas distribution ($\beta > 0.4$) over-fits the datasets (reduced-$\chi^2 < 1$).
Therefore, we fix $\beta=0.4$ for the rest of the analysis.

In Figure \ref{fig-mcmc}, we show the one and two dimensional projections of model parameters'
posterior probability distribution and their 68\% confidence limit (CL) contours. 
The best-fitting values of the model parameters are represented by the mean of posterior 
distribution, whereas their uncertainties are represented by the one dimensional 68\% CL
(see Table \ref{tab-szxr}).

For the power-law model, we obtain $\alpha_{\rm fg}=0.24\pm0.061$
which translates to \fg$\sim$ $3.2^{+1.7} _{-1.1}$\% and $5.5^{+1.8} _{-1.4}$\% (i.e. a baryon budget of 
$\sim$ 20\% and 30\%) 
for virial masses $M_v=10^{12}$ and $ 10^{13} M_{\odot}$, respectively.
The hot CGM fraction increases to 7.7 (13)\% at $M_v=10^{12}\, (10^{13}) M_{\odot}$ i.e. 
a baryon budget of 46 (78)\%, if the same gas density profile is extrapolated out to $2R_v$.
The best-fitting value of $\alpha_{T}=-0.59^{+0.071} _{-0.12}$ compensates for the increasing virial temperature
with virial mass, giving $\rm T_{gas} \sim 0.21^{+0.04} _{-0.05}$ keV at $ M_v \sim 10^{13} M_{\odot}$.
The constraints on $\alpha_{\rm fg}$ and hence hot gas fraction are driven by the X-ray measurements as 
the X-ray luminosity is highly sensitive to the underlying gas distribution. However, X-ray emission weekly
depends on the gas temperature thus giving poor constraints on $\alpha_{T}$. On the other hand,
tSZ is degenerate between gas density and temperature. Combining tSZ with X-ray breaks this degeneracy 
and the constraints on $\alpha_{T}$ are primarily driven by tSZ, which favours a lower gas temperature
as both hot and warm gas contribute to the tSZ signal. 

\subsection{Isothermal hydrostatic equilibrium}
Next, we explore an isothermal distribution of the hot CGM in hydrostatic equilibrium with 
the dark matter halo with the gas density profile given by,
\begin{equation}
 n_e(r) \propto \exp\Bigl[ -\Bigl(\frac{\mu m_p G M_v}{k_b T_{\rm gas} R_s}\Bigr) \frac{1-\log(1+r/R_s)/(r/R_s)}{\log(1+C_v)
 -C_v/(1+C_v)}\Bigr]
 \label{eqn-hydro}
\end{equation}
where $R_s (\equiv R_v/C_v)$ is the scale radius and $C_v$ is the concentration parameter of the
dark matter halo \citep{duffy08}. The hot gas fraction and the temperature are
determined by Equation \ref{eqn-tv}. 

A hydrostatic model (reduced-$\chi^2 \approx 0.64$)
prefers a higher value of $\alpha_{\rm fg}$ and hence a lower gas fraction.
The best-fitting value of $\alpha_{\rm fg} \sim 0.48^{+0.027} _{-0.051}$ predicts \fg $\sim$ $0.6^{+0.3} _{-0.1}$\% and 
$1.8^{+0.5} _{-0.2}$\% (i.e. a baryon budget of 4\% and 11\%) for the 
virial masses $M_v=10^{12}$ and $ 10^{13} M_{\odot}$, respectively.
At the same time, the gas temperatures are higher (i.e. a lower value of $\alpha_T$) than 
a simple power-law gas distribution. The best-fitting value of $\alpha_T \sim -0.33 ^{+0.052} _{-0.023}$ gives 
$T_{\rm gas} \sim 0.38^{+0.05} _{-0.02}$ keV at $M_v=10^{13} M_{\odot}$. The main difference between the power-law and 
the hydrostatic equilibrium model is that the temperature of the hydrostatic model directly affects its  
gas density profile (see Equation \ref{eqn-hydro}). Hydrostatic equilibrium tries to keep the gas temperature
close to the virial temperature and a higher temperature leads to a lower gas fraction.
Extrapolating the density profile out to $2R_v$ gives \fg $\sim$ 2.2 (4)\% at 
$M_v=10^{12}\,(10^{13}) M_{\odot}$ i.e. a baryon budget of $\sim$ 13 (24)\%.

In Figure \ref{fig-uncert}, we compare the stacked CGM X-ray luminosity and the tSZ measurements with 
the predictions of our best-fitting models along with their 68\% uncertainty region.
The power-law model predicts a larger X-ray and tSZ signal throughout the mass range considered except near 
the upper mass end where the hydrostatic model predicts larger X-ray luminosities. 
The power-law model also allows a larger uncertainty in the predicted signal owing to the larger 
uncertainties in the model parameters (see Table \ref{tab-szxr}).

\section{Discussion}
\subsection{Comparison with the hot halo of the Milky Way}
The hot halo of the Milky Way has been studied in detail through a variety of methods (see \citealt{bland16} 
for a recent review of the Milky Way observations).
The best estimate of the virial mass of the Milky Way is $\sim 1.3 \times 10^{12} M_{\odot}$ which
translates to $M_{500} \sim 8.2 \times 10^{11} M_{\odot}$.
Therefore, the Milky Way lies below the lowest mass tSZ/X-ray data points used for our analysis.
In Figure \ref{fig-mw}, we extrapolate our best-fitting predictions for $M_v=1.3\times 10^{12} M_{\odot}$  
(at $r>0.15\times R_{500}$) and compare it with the observed density of the hot halo of the Milky Way from
the following;
\begin{enumerate}
 \item The hot CGM density required to explain the observed ram pressure stripping of the dwarf
satellites of the Milky Way from \citep{putman09};
\item An adiabatic hot halo (assuming the hot gas to contain 10\% of the total halo mass, 
$M_v=10^{12} M_{\odot}$), shown to be consistent with a number of independent observations \citep{fang13};
\item CGM profile derived from {\it XMM-Newton} observations of OVII and OVIII emission
lines along $\sim650$ sightlines,and OVII and OVIII absorption
lines in the background quasar spectra, assuming a power-law model ($n_e \propto r^{-3\beta}$) \citep{miller15};
\item Results of a 3-D hydrodynamic simulations
including stellar feedback, radiative cooling and the cosmological accretion for a 
Milky Way type galaxy with $M_{200}=10^{12} M_{\odot}$ \citep{fielding17}. 
\end{enumerate}
Note that, \cite{miller15} start with a $\beta$-model and core radius 
$\sim 5$ kpc, and then approximate it to a simple power-law form since only 4 out of 
649 OVII and OVIII emission lines pass through $r<5$ kpc.
Our best-fitting power-law model gives a hot gas mass $\sim 4.4^{+2.2} _{-1.5} \times 10^{10} M_{\odot}$
($\sim$ 3.4\% of $\rm M_v$), whereas the hydrostatic model predicts $\sim 9.0^{+3.6} _{-1.5} \times 10^{9} M_{\odot}$,
($\sim$ 0.7\% of $\rm M_v$) within 285 kpc.  
The hot gas mass predicted by our power-law model is much smaller 
than the value ($\sim 10^{11} M_{\odot}$ for $M_v=10^{12} M_{\odot}$) assumed by \cite{fang13}.
The hot CGM mass predicted by the $\beta$-model of \cite{miller15} is
$\sim 6\times 10^{10} M_{\odot}$ (within 285 kpc).
\cite{fielding17} find that the CGM density distribution for $M_{200}\sim10^{12}$ M$_\odot$ 
haloes is nearly independent of the variation in stellar feedback.
Using their CGM density profiles, we find that the 
CGM mass is $\sim 3.7\times 10^{10} M_{\odot}$, within 285 kpc.
These mass estimates are within the 68\% CL of our power-law model, whereas our hydrostatic
model predicts a much smaller hot CGM reservoir.

\begin{figure}
\vspace{-2mm}
% \begin{center}
% \includegraphics[width=8.8cm,angle=0.0 ]{nelc-mw.eps}
\includegraphics[width=8.9cm]{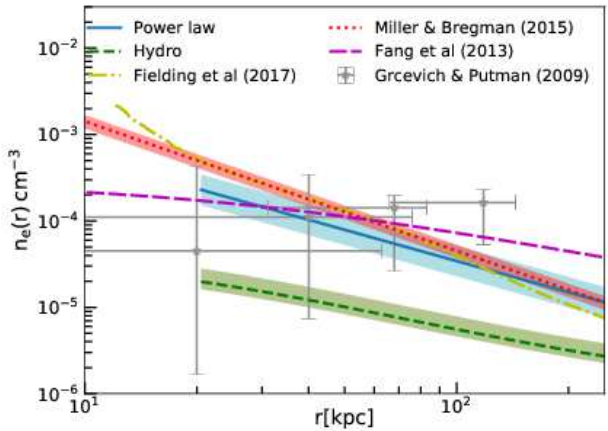}
% \end{center}
\caption{Hot CGM density profile for Milky Way type galaxy for the power-law model (blue solid line), 
the hydrostatic halo model (green dashed line), predictions by Fielding et al 2017 (dot-dashed yellow line) and 
Fang et al 2013 (long dashed magenta line), observations by Miller \& Bregman 2015 (dotted red line) 
and Grcevich \& Putman 2009 (gray points with error bars).
The shaded regions represent the 68\% CL
in the electron density determination.}
\label{fig-mw}
\end{figure}

\subsection{Comparison with other phases of CGM}
The  cool, warm and the hot CGM are the three main phases of CGM that contribute significantly 
to the baryon census of the galaxy (see \cite{tumlinson17}, their section 5.2, for a recent compilation).
The mass budget of the cool CGM, best studied through low ions (CII, CIII, SiII, SiIII, MgII etc.)
in the UV absorption line spectroscopy, is around 
$10^{10}-10^{11} M_{\odot}$ for low redshift $L^*$ galaxies \citep{werk14, stern16, prochaska17}.
The mass budget of the warm CGM, traced by high ions (CIV, OIV etc.) has higher
uncertainty due to the uncertainty in the ionization mechanism. For $L^*$ galaxies, the warm CGM 
contains $>2\times 10^9 M_{\odot}$ \citep{tumlinson11}. 
Our hot CGM mass estimate ($\sim 6\times 10^{9}-7.5\times 10^{10} M_{\odot}$ for the two models at 
$M_v\sim 1-2 \times 10^{12}$ M$_\odot$) is comparable to or larger than the mass contained in other CGM phases.
Together, the cool, warm and the hot phases of CGM offer a potential solution to the galactic missing baryon problem. 

In Figure \ref{fig-surf}, we compare the surface density profile
of the hot CGM (at $M_v=1.3\times 10^{12} M_{\odot}$ 
integrated out to $2R_v$) with the profiles of other CGM phases (from Figure 7 of \citealt{tumlinson17}).
Our power-law model for the hot component dominates the surface density at $r/R_v > 0.2$, whereas,
the hydrostatic hot halo model is subdominant at all radii.

\begin{figure}
\vspace{4mm}
% \begin{center}
% \includegraphics[width=8.8cm,angle=0.0 ]{nelc-mw.eps}
\includegraphics[width=8.2cm]{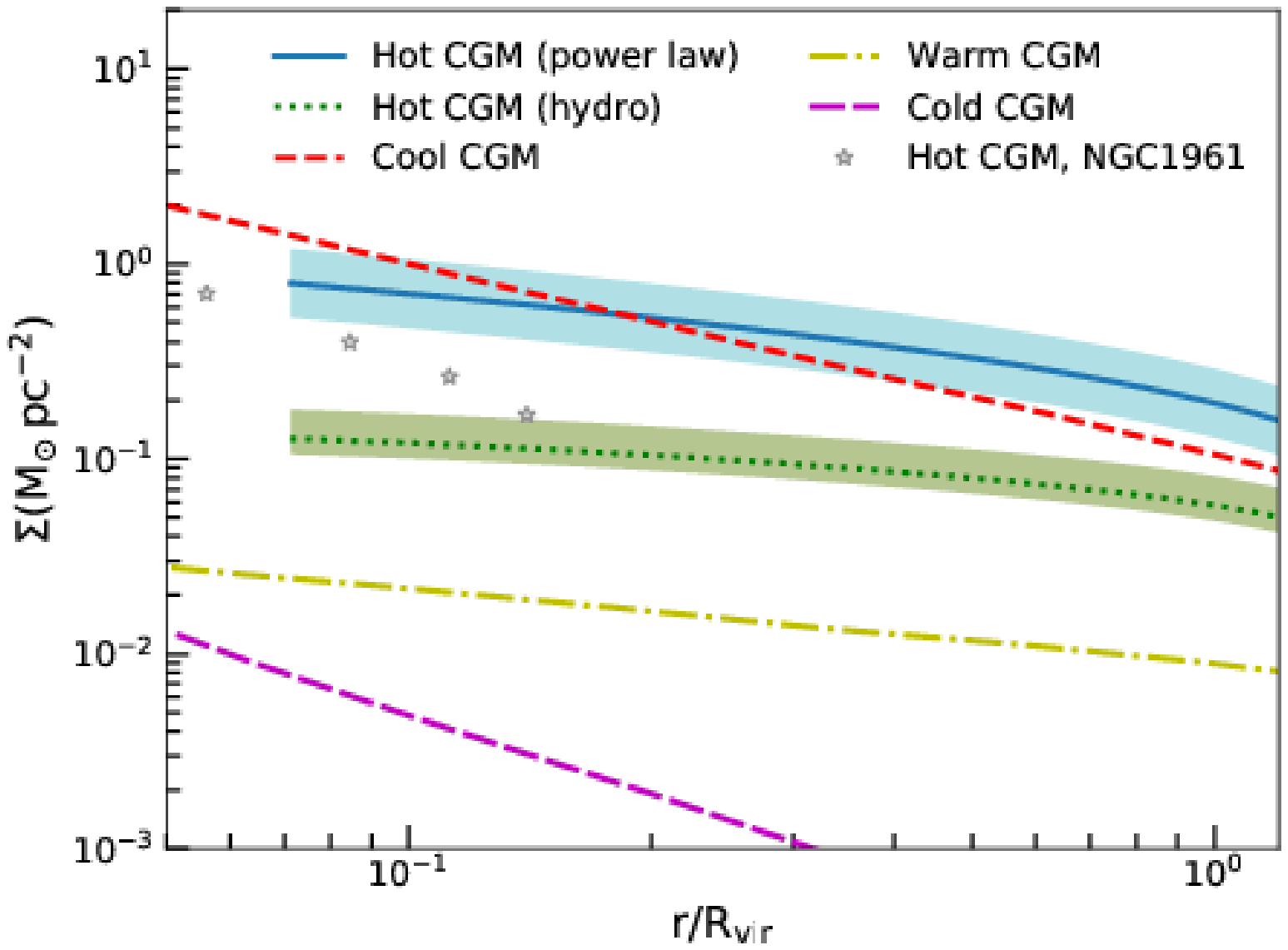}
% \end{center}
\caption{Surface density profile of the hot CGM for the power-law model (solid blue line) and the hydrostatic halo 
model (dotted green line), cool CGM (dashed red line, \protect\citealt{werk14}), warm CGM (dot-dashed yellow 
line, \protect\citealt{tumlinson11, peeples14}), cold CGM 
(long dashed magenta line, \protect\citealt{zhu13}) and the hot CGM in NGC 1961 (gray points, \protect\citealt{anderson16}).
Note that beyond $\approx 0.2 R_{\rm vir}$, the hot CGM component in case of our power law model
dominates over all the other phases of the CGM.}

\label{fig-surf}
\end{figure}

\subsection{Impact of uncertainties}
\label{sec-error}
In this section, we estimate the robustness of our results to the uncertainties surrounding the fiducial parameters and
observed signal:

%\subsubsection{Uncertainty in hot CGM temperature and metallicity}
%\label{sec-temp-met}
(i) Uncertaintes in the hot CGM temperature and metallicity - Our temperature profiles are normalized to give 
 $f_{12}=3.4$ i.e.
$T_{\rm norm} = 2\times10^6$ K for a $10^{12} M_{\odot}$ galaxy. 
These choices are motivated by the observations of the Milky Way 
and external galaxies with $M_v \gtrsim 10^{12} M_{\odot}$ 
(using spectral analysis of the inner halo $\lesssim 0.1$-$0.2 R_v$
in case of external galaxies).  
Given the large scatter in the properties for individual spiral galaxies \citep{li17, anderson13},
we explore the impact of our choice of temperature and metallicity by changing the temperature normalization 
to $T_{\rm norm} \rightarrow T_{\rm norm}/2$ ($f_{12}=1.7$)
and $2\times T_{\rm norm}$ ($f_{12}=6.8$). Increasing (decreasing) the 
temperature normalization by a factor of two, results in $\alpha_{\rm fg}$ changing by $\sim 8\%$, which 
leads to decrease (increase) of \fg by 12.5\% (15\%), respectively, for $10^{12} M_{\odot}$ haloes and a decrease 
(increase) of \fg by 8\% (9\%), respectively, for $10^{13} M_{\odot}$ haloes. The corresponding change in
$\alpha_T$ is $\sim 33\%$, resulting in the change 
in the CGM temperature by roughly 30\%, either way for $10^{13} M_{\odot}$ haloes.
These effects are even smaller for the hydrostatic model.
Instead of a fixed $f_{12}$, we also check to see its impact if
left free as one extra parameter reflecting any unknown uncertainty in the
temperature normalization. We find that using $f_{12}$ as a free parameter with the 
reasonable uniform prior in the range $[1.7-6.8]$,  i.e. a 100\% uncertainty in
our fiducial normalization, has negligible effects on the resultant gas fraction and temperature.

We have, further, used a fixed metallicity of $0.2 Z_{\odot}$ for all masses. Increasing 
$Z=0.2$ to $0.4 Z{\odot}$ has negligible effects on CGM temperature estimation, whereas, there is a  
small decrease in the hot CGM fraction by a factor 1.5 for the power-law and 1.2 for the hydrostatic model.

%\subsubsection{Uncertainty in the stacked tSZ/X-ray signal}
%\label{sec-2h}
(ii) Uncertainty in the stacked tSZ/X-ray signal - Our analysis assumes that P13 and A15 
measurements represent the true one-halo term.
However, a recent study by \cite{vikram17} points out that the two-halo term in SZ-group cross-correlation 
function can dominate over the one-halo term 
in  $M_v \lesssim 10^{13} M_{\odot}$ haloes.
P13 avoid the two-halo contribution to $Y_{500}$ by applying certain 
isolation criteria (see P13 for details)
to their galaxy catalogue \footnote{However, \cite{hill17} argue that the isolation criteria used by P13
are not robust for
low mass haloes.}. The stacked tSZ signal is also marginally detected at halo masses $M_{500}<10^{13} M_{\odot}$.
%Another study by \cite{greco15} claim that the stacked signal in this low mass regime may be completely due to dust and not SZ. 
\cite{wang16} used almost the same sample of LBGs to measure the stacked
weak gravitational lensing signal. Their estimated effective halo mass, for a given stellar mass bin, are lower
compared to P13/A15 results. This shift
in halo mass is equivalent to 30\% increase in tSZ signal and 40\% increase in X-ray signal, which can lead to a
larger hot CGM content. 
%They also compute the uncertainty in the effective halo mass, not available in P13 and A15.

Incorporating the above uncertainties in our analysis is beyond the scope of this paper. 
However, to bypass our lack of understanding of the tSZ signal, we estimate the hot CGM content using X-ray data only.
For the power-law density profile, X-ray emission alone gives
$\alpha_{\rm fg}=0.29^{+0.07} _{-0.03}$ and $\alpha_{T}=-0.1\pm0.6$. While $\alpha_{T}$ is unconstrained as
expected, the constraint on $\alpha_{\rm fg}$ agrees with the joint X-ray-tSZ 
constraint to within $1\sigma$.
For the hydrostatic model, both the parameters are poorly constrained 
($\alpha_{\rm fg}=0.34^{+0.30} _{-0.27}$,  $\alpha_{T}=-0.11\pm0.38$) since the 
temperature uncertainties feed into the density estimation. 

\section{Summary}
We have obtained the joint X-ray-tSZ constraints on the hot CGM mass fraction and temperature
for massive galaxies. The datasets used in this paper are from P13 (stacked tSZ) and A15 (stacked X-ray luminosity).
The two CGM density profiles considered here, namely the power-law and hydrostatic halo 
model are based on the assumptions that, 1) the gas is isothermal, 2) the gas temperature at $M_v=10^{12} M_{\odot}$ is 
$\sim 2\times 10^6$ K and 3) a uniform metallicity, $0.2 Z_{\odot}$. 
The main conclusions of this work are the following.

\begin{itemize}
\item The power-law model predicts \fg $\sim 3.2^{+1.7} _{-1.1}$\% and
 $5.5^{+1.8} _{-1.4}$\% for the halo masses $10^{12}$ and $10^{13} M_{\odot}$, respectively.
 Therefore, the hot CGM holds approximately 20-30\% of baryonic mass in massive haloes.
 The predicted gas temperature at $M_v=10^{13} M_{\odot}$ is $\sim 0.21^{+0.04} _{-0.05}$ keV, only slightly
 larger than the temperature at $M_v=10^{12} M_{\odot}$ (0.17 keV).
 
\item The hydrostatic halo model predicts lower hot gas fractions (0.6$^{+0.3} _{-0.1}$\%
 and 1.8$^{+0.5} _{-0.2}$\% at $M_v=10^{12}$ and $10^{13} M_{\odot}$,
 respectively) and higher temperatures ($T_{\rm gas} \sim 0.38^{+0.05} _{-0.02}$ keV at $M_v=10^{13} M_{\odot}$)
 as compared to the power-law model. This translates to a baryon budget of 4-11\%.
 
\item Extrapolating the density profiles to $2R_v$ increases the baryon budget in the hot CGM to 46 (78)\%
for the power-law model and 13 (24)\% for the hydrostatic hot halo model at $M_v=10^{12}$ ($10^{13}$) $M_{\odot}$.
Note that, a recent study by \cite{lim17}
stacked the kinetic Sunyaev-Zel’dovich (SZ)
signal from galaxy groups down to the halo mass $\sim 10^{12.3} M_{\odot}$ and showed that their results
are consistent with the CGM containing galactic cosmic baryon fraction in the warm phase ($T_{\rm eff} \sim 10^5-10^6$ K)
within the virial radius of the galaxy, however with large uncertainties in the CGM mass fraction.
They use a $\beta$-profile (with $\beta=0.86$) to extract the signal within $3R_{500}$ and use the same profile to
obtain the signal within $R_{500}$. However, such conversions are sensitive to the assumed density profile (see
section \ref{sec-tsz-data}) and may result in the overestimation of the signal, especially for low mass haloes.
 
\item The predictions of the power-law model (extrapolated to $M_v\sim 1.3\times 10^{12} M_{\odot}$)
 are in agreement with the observations of the Milky Way, whereas the hydrostatic model predicts a low density and 
 hot gas fraction. Our estimate of the hot CGM mass is comparable to or larger than
 the mass predicted in other phases (cool and warm phases)
 of the CGM in $L^*$ galaxies.
 
\item Relaxing our assumptions about the gas temperature and metallicities around their fiducial values 
has only small effects on the best-fitting values of the model parameters, given the uncertainties in these parameters. 

\item It is difficult to explore the variations in $\beta$, the temperature profile and gas metallicity across
the mass range due to the paucity of data in this mass range.
Additionally, the observations of the hot CGM in individual massive galaxies
are limited to 10-20\% of $R_v$. A large fraction of the hot CGM is expected to be distributed
out to the virial radius, making it difficult to directly compare them with our results.
The situation is expected to improve with more observations in future.\\
\end{itemize}
{\bf{ACKNOWLEDGEMENTS}}\\
We thank the anonymous referee for valuable suggestions
and comments.
We thank Michael E. Anderson, Eiichiro Komatsu, Yin-Zhe Ma and Saurabh Singh for helpful discussions. 

\footnotesize{
\bibliography{bibtexcgm}{}
\bibliographystyle{mn2e}
}

\end{document}